\documentclass[10pt,a4paper,english]{article}
\usepackage{amssymb,amsmath,multicol}

\usepackage{cite}
\usepackage[T1]{fontenc}
\usepackage{subfigure}
\usepackage{amssymb}
\usepackage{esint}
\usepackage{a4wide}
\usepackage{psfrag}
\usepackage[dvips]{graphicx}
\usepackage{babel}
\usepackage{color}
\usepackage{hyperref}
\usepackage{graphicx}
\usepackage{dcolumn}
\usepackage{bm}
\usepackage{latexsym,amsmath,amssymb,amsfonts,amsthm,enumerate}
\newcommand{\EQ}{\begin{equation}}
\newcommand{\EN}{\end{equation}}
\newcommand{\bea}{\begin{eqnarray}}
\newcommand{\eea}{\end{eqnarray}}
\renewcommand{\(}{\left (}
\renewcommand{\)}{\right )}
\renewcommand{\[}{\left [}
\renewcommand{\]}{\right ]}

\newcommand{\ra}{\rangle}
\renewcommand{\l}{\lambda}

\newcommand{\dg}{\dagger}
\newcommand{\ep}{\epsilon}
\renewcommand{\=}{\simeq}
\newcommand{\Wt}{\tilde{W}_0}
\newcommand{\nn}{\nonumber}
\newcommand{\ee}[1]{
\begin{align}
#1
\end{align}}
\long\def\symbolfootnote[#1]#2{\begingroup%
\def\thefootnote{\fnsymbol{footnote}}\footnote[#1]{#2}\endgroup}

\newcommand{\eq}[2][]{
\begin{equation}
#2 \label{#1}
\end{equation}} 

\begin{document}


\vspace*{0.5cm}
\begin{center}
\LARGE
Zamolodchikov-Faddeev Algebra and Quantum Quenches \\
in Integrable Field Theories 
\end{center}
\vspace{0.5cm}

\begin{center}
{\large S. Sotiriadis$^{1,2}$, D. Fioretto$^{1,2,}$\symbolfootnote[4]{Current address: Physics Department, University of Fribourg, Chemin du Mus\'ee 3, 1700 Fribourg, Switzerland},  G. Mussardo$^{1,2,3}$
\vspace{0.9cm}}

{\sl $^1$SISSA,
Via Bonomea 265 Trieste, Italy\\[2mm]
$^2$INFN, Sezione di Trieste\\[2mm]
$^3$The Abdus Salam International Centre\\ for Theoretical Physics, Trieste, Italy
}

\end{center}

\vspace{0.5cm}
\begin{center}
{\bf Abstract}\\[5mm]
\end{center}
We analyze quantum quenches in integrable models and in particular the determination of the initial state in the basis of eigenstates of the post-quench hamiltonian. This leads us to consider the set of transformations of creation and annihilation operators that respect the Zamolodchikov-Faddeev algebra satisfied by integrable models. We establish that the Bogoliubov transformations hold only in the case of quantum quenches in free theories. In the most general case of interacting theories, we identify two classes of transformations. The first class induces a change in the $S$-matrix of the theory but not of its ground state, whereas the second class results in a ``dressing'' of the operators. As examples of our approach we consider the transformations associated with a change of the interaction in the Sinh-Gordon and the Lieb-Liniger model.

\date{\today}


\newpage

\section{Introduction}

A quantum quench is an instantaneous change in the parameters that determine the dynamics of an isolated quantum system e.g. the masses or coupling constants of its hamiltonian. This topic has recently attracted a lot of attention as shown by the increasing number of papers addressing this issue (for a recent review, see \cite{Silva} and references therein). From an experimental point of view this is a feasible way to bring the system out-of-equilibrium and study its evolution under the quantum mechanical natural laws, in isolation from the environment. In particular, the scientific interest in quantum quenches started growing after the experimental realization of global sudden changes of the interaction in cold atom systems, a novel technology where quantum statistical physics can be experimentally demonstrated and probed \cite{Greiner,Sadler,Paredes,Kinoshita}. From a theoretical point of view the problem consists in preparing the system in a particular trial state, which is typically the ground state of some hamiltonian, and study its evolution under a different hamiltonian \cite{Barouch_I, Barouch_II,Barouch_III ,Igloi,Sachdev,CC1,CC2,Cazalilla,Cramer2,Cramer1,FCC,MC}. Apart from being one of the simplest and well-posed ways to study out-of-equilibrium quantum physics, quantum quenches also give rise to a fundamental long-standing open question of central importance in statistical physics, the question of thermalization: how do extended quantum physical systems tend to thermal equilibrium starting from an arbitrary initial state?

Of particular interest is the case of (1+1)-dimensions where a discrimination between integrable and non-integrable systems is possible. Integrable models are models that exhibit factorization of the scattering matrix and can be solved exactly (see, for instance, \cite{Mussardo} and references therein). Their classical counterparts possess as many integrals of motion as their degrees of freedom and this fact prevents thermalization of an arbitrary initial state, as not all of the micro-states of equal energy respect the conservation of all other integrals of motion. This property is also expected to hold at the quantum level. In a seminal experiment \cite{Kinoshita} it was observed that a trapped (1+1)d Bose gas, initially prepared in a non-equilibrium state, does not thermalize but tends instead to a nonthermal momentum distribution. The absence of thermalization suggests as a possible reason the integrability of the system which approximates a homogeneous (1+1)d Bose gas with point-like collisional interactions, a typical integrable model, even though the confining potential used in the experiment breaks the homogeneity and therefore integrability of the system. This experiment triggered an intense discussion about the role of non-integrability in the thermalization process. It was soon conjectured \cite{Rigol} that in an integrable case the system does exhibit stationary behavior for long times, described however not by the usual Gibbs ensemble but a Generalized Gibbs Ensemble (GGE) where new Lagrange multipliers are introduced into the density matrix, one for each integral of motion, for accounting their conservation (in the same way that the inverse temperature $\beta$ is the Lagrange multiplier corresponding to the constraint of energy conservation)
\EQ
\rho\,=\, Z^{-1} \exp{\(-\sum_m \lambda_m \mathcal{I}_m\)}\,\,\,.
\label{GGE}
\EN
This conjecture has been shown to be correct in many different special cases, both by analytical and numerical methods \cite{Rigol2,Barthel,Manmana,Rossini,FM, Essler,Iucci,MC}. On the other hand, it has not become yet clear whether non-integrability alone is sufficient to ensure thermalization or not \cite{Banuls,Gogolin} neither has exact thermalization been firmly demonstrated as an outcome of unitary evolution. For instance, recent analysis suggests that the behavior may be more complicated and may depend on the initial state, finite size effects and locality \cite{Kollath,Biroli,BDKM}. For recent experimental developments backed by numerical simulations we refer to \cite{exprm1,exprm2,exprm3}.

Concerning with the analytic approach to the problem, in the paper \cite{FM} it was shown that any quantum quench in an integrable quantum field theory where the initial state has the form 
\EQ
| \psi_0\ra \sim \exp\(\int d\theta \; K(\theta) Z^\dg(\theta) Z^\dg(-\theta)\) | 0 \ra\,\,\,,
\label{squeezed}
\EN 
leads to stationary behavior as described by the GGE ansatz. In the expression above, $| 0 \rangle$ is the vacuum state of the theory, while $Z^\dg(\theta)$ is the creation operator of the quasi-particle excitation, which satisfies the  relativistic dispersion relations $E = m \cosh\theta, P = m \sinh\theta$, where $\theta$ labels the rapidity of the particle of mass $m$. As it is clear from this expression, the function $K(\theta)$ is the amplitude relative to the creation of a pair of particles with equal and opposite rapidity.

The above form of quench state is also called "squeezed coherent" state. The reason to choose such an initial state comes from its relation with boundary integrable states (i.e. boundary states that respect the integrability of the bulk theory) and from the technical advantages it exhibits. It is however true that this requirement is satisfied in general for quantum quenches in a free theory, bosonic or fermionic, as well as for the important cases of Dirichlet and Neumann states in integrable field theory. These states are supposed to capture the universal behaviour of all quantum quenches in integrable models, if renormalization group theory expectations are also applicable out-of-equilibrium. They have been successfully used earlier \cite{CC1,CC2}, applying a Wick rotation from real to imaginary time which allows a mapping of the original quantum quench problem to an equilibrium boundary problem defined on a euclidean slab with boundary conditions both equal to the initial state right after the quench. Although this approach does not help in determining the expression of the initial state as a function of the quench parameters, it has led to correct predictions in certain important asymptotic limits.

It still remains to find out from first principles, whether this assumption for the form of the initial state holds in general for any quantum quench in an integrable system or, if not, under what conditions this happens. Our method to attack this problem begins by understanding the fundamental reason why this condition holds generally for free systems and then investigating if this reason can be generalized to the integrable case. It turns out that in free systems the reason lies in the fact that the relation between the creation-annihilation operators before and after the quench is of linear Bogoliubov type, which itself is a consequence of their canonical commutation or anti-commutation relations. In integrable theories these commutation relations are replaced by the so-called Zamolodchikov-Faddeev (ZF) algebra which, assuming for simplicity that there is only one quasiparticle in the theory, can be written as
\begin{align}
& Z(\theta_1) Z(\theta_2)= S(\theta_1-\theta_2) Z(\theta_2) Z(\theta_1) \nn \\
& Z(\theta_1) Z^\dg(\theta_2)= S(\theta_2-\theta_1) Z^\dg(\theta_2) Z(\theta_1) + \delta(\theta_1-\theta_2) \label{ZF-alg1}
\end{align}
Intuitively this means that the exchange of two quasiparticles is done by the scattering matrix $S(\theta)$. Then a natural question arises: what are the possible transformations of creation-annihilation operators that respect the above algebra? This is a question of more general interest in both the abstract mathematical description of integrable field theories and their potential physical applications in concrete models. In this article we show that, unlike in free theories, the ZF commutation relations do not admit Bogoliubov transformations and we construct several other classes of non-trivial infinitesimal transformations.

We start our presentation by first discussing the structure of initial states in global quantum quenches. Then we outline a general strategy on how to determine the initial state from the relation between the ZF creation-annihilation operators before and after the quench and derive the conditions that must be satisfied by infinitesimal transformations of these operators in order to respect the ZF algebra. After showing that in interacting theories the linear Bogoliubov transformations do not leave the ZF algebra invariant, we find two types of acceptable transformations, the first of which induces a shift in the $S$-matrix but does not affect the ground state of the theory, while the second does not change the $S$-matrix but it does change the ground state and the corresponding transformed ground state is, under certain conditions, of the squeezed coherent state form. We note that these are special classes of transformations and outline how more general ones can be constructed. Next we apply these ideas in two typical integrable models, the Sinh-Gordon model and the Lieb-Liniger model, deriving some examples of infinitesimal transformations of their ZF operators that demonstrate the presence of the first type constructed before. Finally we summarize our findings, giving directions for their application to concrete quantum quench problems. There are also two appendices: in Appendix A we discuss the squeezed states in free quantum field theories while in Appendix B we discuss the derivation of the two classes of generators of the ZF algebra transformations.

\section{On the initial states in global quantum quenches}
A quench process consists of preparing the system in a state $| \psi_0 \rangle$ that is not an eigenstate of its hamiltonian $H$ and let this state unitarily evolve according to $H$. At time $t$,  the expectation values of local observables $\Lambda(r)$ are given by
\EQ
\langle \Lambda(t,r) \rangle \,=\,\langle \psi_0 | e^{i H t} \,\Lambda(r)\, e^{-i H t} | 
\psi_0 \rangle \,\,\,,
\EN 
with similar expressions for higher point correlation functions. As evident from the expression above, important information on the subsequence dynamics of the system is encoded in the initial state $| \psi_0 \rangle$. Relevant features of this state can be derived on the basis of general considerations for extended quantum systems having particle excitations and for global quenches. First of all, by relativistic (or even galilean) invariance, we can always assume that the quench state $| \psi_0 \rangle$ carries no momentum. Let $Z^{\dagger}(p)$ be the creation operator of a particle excitation\footnote{To simplify the following formulas we assume that the system has only one neutral elementary excitation, with normalization proportional to the $\delta$ function, $\langle Z(p_1)  Z^{\dagger}(p_2) \rangle \propto \delta(p_1-p_2)$.} of the system of momentum $p$ and let us assume that a basis of the Hilbert space is given by the multi-particle excitations, eigenvectors of the hamiltonian $H$. Then the general form of the initial state $| \psi_0\rangle$ for a global quench is given by an {\em infinite} superposition of multi-particle states of zero momentum 
\EQ
| \psi_0 \rangle \,=\,\sum_{n=0}^{\infty} \int dp_1\,\ldots\, dp_n \tilde{\mathcal K}_n(p_1,\ldots ,p_n) \delta(\sum_{i=1}^n p_i) Z^{\dagger}(p_1) \cdots Z^{\dagger}(p_n) | 0 \rangle \,\,\,,
\label{infinitesup}
\EN 
where $| 0 \,\rangle$ is the vacuum state of the system. This requirement is due to a thermodynamics argument related to the formulation of quench dynamics in $d$ dimensions to the thermodynamics of a $(d + 1)$ dimensional field theory in a slab geometry, where the initial state $|\psi_0 \ra$ plays the role of boundary conditions on both borders of the slab \cite{CC2}. 
In this interpretation of the quench process, the quantity 
\EQ
{\mathcal Z}_0(\tau)\,=\, \langle \psi_0 | e^{-\tau H} | \psi_0 \rangle \equiv e^{-{\mathcal F}_0(\tau)}
\EN 
plays the role of the partition function of the system with boundary conditions fixed by $| \psi_0 \rangle$. For global quenches, the corresponding free energy ${\mathcal F}_0(\tau)$ must be an extensive quantity, ${\mathcal F}_0(\tau) \simeq V f_0(\tau)$, where $V$ is the volume of the system. On the other hand, this quantity can be computed by employing the expression (\ref{infinitesup}) of the initial state $| \psi_0 \rangle$ and a proper normalization\footnote{See \cite{LeClair} for an explicit regularization of this term in the one-dimensional case.} of $\delta(0)$: then the only way to have an extensive behavior in the volume $V$ of the system for ${\mathcal F}_0(\tau)$ is by $| \psi_0 \rangle$ containing an inﬁnite number of multi-particle states.

Notice that one way to automatically take into account the zero momentum condition of the initial state $| \psi_0 \rangle$ is to assume that its infinite superposition is made of pairs of particles of equal and opposite momentum, i.e. ``Cooper pairs''
\EQ
| \psi_0 \rangle \,=\,\sum_{n=0}^{\infty} \int dp_1\,\ldots dp_n {\mathcal K}_{2n}(p_1,\ldots ,p_n) Z^{\dagger}(-p_1) Z^{\dagger}(p_1) \cdots Z^{\dagger}(-p_n) Z^{\dagger}(p_n) |0 \rangle \,\,\,.
\label{infinitepairs}
\EN  
It should be stressed, though, that this formula is a particular case of the more general form (\ref{infinitesup}). But even with this simplification, to specify the initial state $|\psi_0 \rangle$ one still needs an infinite number of amplitudes ${\mathcal K}_{2n}(p_1,\ldots,p_n)$. The great technical advantage of the squeezed coherent states, whose concise expression is given by 
\EQ
| \psi_0 \rangle \,\sim
\exp\(\int dp \; K(p) Z^\dg(p) Z^\dg(-p)\) | 0 \ra\,\,\,,
\label{squeezed1}
\EN 
becomes then evident. In this case, in fact, all the multi-particle amplitudes ${\mathcal K}_{2n}(p_1,\ldots,p_n)$ can be expressed in terms of products of the single amplitude $K(p)$ entering (\ref{squeezed1}), therefore greatly simplifying the problem. 

Squeezed coherent states naturally appear in two contexts: (i) in the purely boundary integrable field theories considered by Ghoshal and Zamolodchikov \cite{Ghoshal}, where the amplitude $K(p)$ also satisfies additional conditions (boundary unitarity and crossing symmetry) and (ii) in quench processes in free theories, both bosonic and fermionic. In the latter case, it is 
worth noticing that the commutation or anti-commutation relations of the annihilation and creation operators $Z(p)$ and $Z^{\dagger}(p)$ of these theories can be cast in the form of ZF algebra (\ref{ZF-alg1}) with $S = 1$ for the boson and $S= -1$ for the fermion. The only parameter entering these theories is in this case the mass of the their excitation and, as shown in detail in Appendix A, its sudden change can be taken into account by a Bogoliubov transformation of the annihilation and creation operators. 
Since the Bogoliubov transformations leave the commutation or anti-commutation relations invariant, in turn they can be seen as the
transformations which leave invariant the ZF algebra of free theories. 
This observation leads us to investigate a more general class of transformations of the ZF operators in interacting integrable field theories which leave their algebra invariant.

\section{Quenches in Integrable Systems}\label{sec_qq}

In this section we analyze the quantum quenches in systems which are integrable before and after the sudden change of one parameter $Q$, which can be for instance the mass of the particle or the coupling constant of the theory. One of the main issues of this problem is to write down the pre-quench state (usually the vacuum, annihilated by the pre-quench particle operators) in terms of the post-quench particle basis. This task involves in principle the computation of an infinite number of inner products, an operation usually difficult to fulfill (for a discussion of related numerical issues see for example \cite{FCC,Gritsev}). Therefore, it would be useful to have a different approach. In principle, a possible way to determine the initial state $| \psi_0 \rangle $ in terms of the post-quench creation-annihilation operators $Z,Z^\dg$ is to implement the following program:
\begin{itemize}
\item for an arbitrary value of the parameter $Q$, find initially the relation between the ZF $Z$ operators of the theory and the physical field operator $\phi$, i.e. $\phi=f(Q;Z_Q)$. 
\item use the continuity of the field as boundary condition in the quench process $Q_0  \rightarrow Q$  
\eq{f(Q_0;Z_{Q_0})=f(Q;Z_Q)}
for deriving the relation between the old and the new ZF operators
\eq[eq1]{Z_{Q_0}=F(Q_0,Q;Z_{Q})=f^{-1}(Q_0;f(Q;Z_Q))}
\item write the initial state $|\Omega_0\ra$, which is known in the pre-quench ZF basis (and is typically the ground state defined by $Z_{Q_0}|\Omega_0\ra=0$), in the new basis using the above relation. 
\end{itemize}
If this program can be realized, the time evolution of the initial state in the new basis can be  computed easily. Going in more detail, the first step of this program consists in expanding the physical field operator as a series in the ZF operators using all of its form factors, i.e. the matrix elements of the field $\phi(x)$ in the asymptotic states. The second step involves the inversion of this series; this might require an ingenious ansatz for the function $F$. The third step requires to deal with the most general expansion of a state in the post-quench basis and to determine the coefficients of $|\Omega_0\ra$ term by term from the equation $F(Q_0,Q;Z_{Q})|\Omega_0\ra=0$.

While the first step is essentially a re-expression of the body of information obtained by the form factors program, the other steps are in general highly nontrivial. In order to partially circumvent these difficulties, in the following we will exploit some general properties of integrable field theories. As previously said, in free theories the relation between the new and the old creation/annihilation operators is Bogoliubov-type, fixed by the condition of leaving invariant the (trivial) ZF algebra of these theories. Analogously, for generic integrable theories, the transformation between the pre and the post-quench ZF operators must respect the algebra. This leads us to investigate under which condition this requirement is satisfied. Of course this is quite an abstract point of view:  knowing that a certain transformation respects the algebra does not necessarily clarify the physical nature of the quench protocol. Nevertheless, it is surely important to understand what are the possible algebra-preserving transformations and if their form is restrictive enough to make prediction about the initial state.  In Section \ref{sec_physical}, we integrate this analysis with a perturbative study of a typical integrable model, the Sinh-Gordon model, and its non-relativistic counter-part, i.e. the Lieb-Liniger model.

\subsection{Conditions required for transformations of the ZF algebra operators }\label{sub_conditions}

We are looking for transformations of the creation-annihilation operators $Z,Z^\dagger$ that respect the ZF algebra. We also require that the transformations respect the translational invariance of the theory, since we are considering only homogeneous  systems, both before and after the quench. For this reason we will write the ZF algebra in momentum representation which, as we will see soon, serves better this requirement
\begin{align}
& Z_{p_1} Z_{p_2}= S(p_1,p_2) Z_{p_2} Z_{p_1} \\
& Z_{p_1} Z^\dg_{p_2}= S(p_2,p_1) Z^\dg_{p_2} Z_{p_1} + \delta(p_1 - p_2)\,\,\,, 
\end{align}
along with the standard properties of the $S$-matrix
\eq{S(p_1,p_2)^{-1} = S(p_1,p_2)^{*} = S(p_2,p_1) = S(-p_1,-p_2)\,\,\,.}
Notice that in comparison with the form of the ZF algebra in the rapidity representation (\ref{ZF-alg1}), we have redefined\footnote{This definition is also tailored to our purposes, since the energy of particles may change under a quantum quench and we would like to absorb all changes into the transformation of the operators. In addition it is consistent with the usual normalization of energy eigenstates in the free limit which will be useful later.}
 the operators as $Z_p \equiv Z(p) = Z(\theta(p))/\sqrt{E(p)}$ since $\delta(\theta(p)) = E(p) \delta(p)$. 

We focus our attention on infinitesimal transformations, assuming that a finite transformation can be build up by repetitive action of the infinitesimal ones. We also allow for infinitesimal changes of the $S$-matrix\footnote{Note that this assumption may exclude the special case of free bosons. This is because for all integrable field theories except for free bosons, the $S$-matrix at zero momentum is $S(0)\equiv S(p,p)=-1$. Therefore the transition from a free bosonic point of an integrable theory to another point that does not correspond to free bosons is always discontinuous as far as the $S$-matrix is concerned. In the following we sometimes make use of the property $S(0)=-1$ in which case we mention it explicitly.}. Since we demand that the transformations commute with the momentum operator, the new operator must carry the same momentum as the old one but not necessarily the same rapidity, as the quench may involve a change of the mass of particles (this is the reason why the momentum representation suits better our problem). Therefore both the transformed operator and the $S$-matrix are in general expressed as 
\begin{align}
& Z'_{p} = Z_{p} + \epsilon W_{p} \\
& S'(p_1,p_2) = S(p_1,p_2) + \epsilon T(p_1,p_2)
\end{align}
where $\epsilon$ is a small quantity, function of the infinitesimal change $\delta Q$ of the quench parameter. In order to satisfy the ZF algebra, they must fulfill the conditions
\begin{align}
& W_{p_1} Z_{p_2} + Z_{p_1} W_{p_2} = T(p_1,p_2) Z_{p_2}Z_{p_1} + S(p_1,p_2)(Z_{p_2}W_{p_1} + W_{p_2} Z_{p_1}) \label{1} \\
& W_{p_1} Z^\dg_{p_2} + Z_{p_1} W^\dg_{p_2} = T(p_2,p_1) Z^\dg_{p_2}Z_{p_1} + S(p_2,p_1)(Z^\dg_{p_2}W_{p_1} + W^\dg_{p_2} Z_{p_1}) \label{2}
\end{align}
for all values of $p,p'$, along with the following conditions for $T$
\eq[3a]{ T(p_1,p_2)^* = T(p_2,p_1) = T(-p_1,-p_2) = - T(p_1,p_2) S^{-2}(p_1,p_2)}
coming from the unitarity of the $S$-matrix.

The operator $W_p$ can generally be written as an expansion in the operators $Z,Z^\dg$
\eq[gen_sol]{W_p = \sum_{n,m=1}^\infty  \delta\(p + \sum_{i=0}^n q_i - \sum_{j=0}^m p_j \) \; a_{n,m}(\{q_i\},\{p_j\}) \; \prod_{i=0}^n Z^\dg_{q_i} \prod_{j=0}^m Z_{p_j} }
The above conditions are then translated into a sequence of relations between the coefficients $a_{n,m}$ of different orders. Below we construct and study several simple classes of solutions in which the above expansion terminates after a few terms.

\subsection{A first trial: linear Bogoliubov transformations}

Let us initially assume that $W$ corresponds to a linear Bogoliubov transformation which, in the infinitesimal form, means $W_p = a_p Z^\dg_{-p}$. In this case it is easy to see that  
the previous conditions become 
\begin{align*}
S(-p,p') & = S(p,-p') = S(p',p) \\
a_p/a_{-p} & = S(p,p) \equiv S(0)  \\
T(p,p') & = 0
\end{align*}
for all values of $p,p'$. The first of these equations implies that $S(p,p')^2=1$, i.e. $S(p,p') =\pm 1$. We therefore arrive at the interesting result that \emph{the linear Bogoliubov transformation is a symmetry of the algebra only in the trivial case of free fields, bosons or fermions}. Moreover, it is easy to show that any other linear combination of the operators is inconsistent with the general conditions (\ref{1}) and (\ref{2}).

\subsection{Generators of S-matrix changes}

We will now construct a transformation that induces a non-zero change $T$ in the $S$-matrix and show that this transformation is unique, in the sense that any infinitesimal transformation that has the same effect must necessarily involve this one. First, observe that, since linear transformations are already excluded, the $TZZ$ term in (\ref{1}) can only be produced as a $\delta$-function by-product of the commutation of higher order terms in $W$. More preciselly,  $W$ must be of 3rd order and must contain one $Z^\dg$ operator and two $Z$'s, so that the commutation of $W_p$ with $Z_{p'}$ produces a $ZZ\delta $ term. Furthermore, from eq.\,(\ref{1}) we see that the two $Z$ operators in the residual $ZZ\delta$ term, which come originally from $W_p$, must carry momenta $p,p'$ (the same as $W_p$ and $Z_{p'}$) and therefore the $Z^\dg$ operator in $W_p$ must carry momentum $p'$ to ensure that $W$'s total momentum is $p$. Thus $W_p$ is of the form $Z^\dg_{p'} Z_{p'} Z_p$. But this must be true for arbitrary $p'$, so $W$ should necessarily consist of a linear combination of all such terms. All this leads to the ansatz 
\eq[ex1]{W_p = \( \sum_q \alpha_{p,q} Z^\dg_q Z_q \) Z_p\,\,\,.}
Let us verify it explicitly by substituting into the required conditions (\ref{1}) and (\ref{2}), the first of which gives
\eq{T(p_1,p_2) = S(p_1,p_2) (\alpha_{p_2,p_1} - \alpha_{p_1,p_2})\,\,\,,}
while the second 
\eq{T(p_2,p_1) = S(p_2,p_1) (\alpha^*_{p_2,p_1} + \alpha_{p_1,p_2})\,\,\,,}
together with 
\eq[gs1]{\alpha_{p,q} + \alpha^*_{p,q} = 0\,\,\,.}
Remarkably all of these requirements are simultaneously satisfied as long as it holds the condition (\ref{gs1}), i.e. if the coefficients $\alpha_{p,q}$ are purely imaginary. Hence, from now on we set $\alpha_{p,q}=ia_{p,q}$ where $a_{p,q}$ are real functions. Notice that this solution also ensures that the $S$-matrix remains unitary, as expressed by the conditions
(\ref{3a}) for $T$.

Studying in more detail these transformations, one realizes that
\eq[cch]{{Z'}^\dg_{p} {Z'}_{p} = {Z}^\dg_{p} {Z}_{p} + i \ep {Z}^\dg_{p} \[ \sum_q (a_{p,q} + a^*_{p,q}) Z^\dg_q Z_q \] {Z}_{p} = {Z}^\dg_{p} {Z}_{p}}
i.e. the conserved charges
\eq{\hat{\mathcal{Q}}_s = q_s \int d\theta \; e^{s \theta} {Z}^\dg_{\theta} {Z}_{\theta}}
remain invariant (unless the factors $q_s$ depend explicitly on the physical parameters whose infinitesimal change leads to this transformation). This allows us to easily derive the corresponding finite 
transformation\footnote{$\mathcal{P}\{...\}$ denotes a path-ordering integration and $s$ is a continuous parameter along some path.}
\eq{Z'_p = \mathcal{P}\{ \exp\( i \int \hat{\mathcal{I}}_p(s) ds \)\} Z_p \qquad \text{ where } \quad \hat{\mathcal{I}}_p(s) = \sum_q a_{p,q}(s) Z^\dg_q Z_q }
which changes the $S$-matrix as
\eq{S'(p,q) = \exp\[ i \int \(a_{q,p}(s)-a_{p,q}(s)\) ds \] S(p,q) \,\,\,.}

Even though we have constructed this infinitesimal transformation heuristically, it is easy to show that this transformation is the only one that changes the $S$-matrix. Any other transformation that changes the $S$-matrix must necessarily be a linear combination of this one along with some other part that does not change it. Indeed if there was another transformation $W'$ that also shifts $S$ to the same $S+\epsilon T$ then from (\ref{1}) and (\ref{2}) their difference $W-W'$ would not change the $S$-matrix. We can therefore decompose any transformation that respects the ZF algebra (infinitesimal or finite) into two parts, one of which is of the above form and performs the shift of the $S$-matrix to the desired value, while the rest leaves it invariant. In this way we have reduced the problem of finding the symmetries of the ZF algebra to the task of identifying those transformations which do not alter the $S$-matrix and which satisfy (\ref{1}) and (\ref{2}) with $T=0$, i.e. 
\begin{align}
& W_{p_1} Z_{p_2} + Z_{p_1} W_{p_2} = S(p_1,p_2)(Z_{p_2}W_{p_1} + W_{p_2} Z_{p_1}) \label{1.1} \\
& W_{p_1} Z^\dg_{p_2} + Z_{p_1} W^\dg_{p_2} = S(p_2,p_1)(Z^\dg_{p_2}W_{p_1} + W^\dg_{p_2} Z_{p_1}) \label{2.1}
\end{align}

\subsection{Other classes of transformations}\label{sec1}

Having reduced the problem to identifying the transformations of the ZF operator which do not alter the $S$-matrix, we will now consider more general classes of symmetry transformations of the ZF algebra. Let us assume initially that $W$ is simply a single product of $Z, Z^\dg$ operators. In order to check the condition (\ref{1.1}), we have to consider how $W_p$ commutes with $Z_{p'}$: when we swap $Z_{p'}$ with each of the operators in $W$ one-by-one, this operation gives as output, for each of these terms, multiplicative $S$-matrix factors as well as additive $\delta$-functions for each $Z^\dg$, which are lower order products. One obvious way to satisfy (\ref{1.1}) is then to choose $W_p$ in such a way that (a) the overall $S$-matrix factor is simply equal to $S(p,p')$ and (b) that the residual lower order terms vanish. 

Let us firstly focus our discussion on the point (a). We assume that $W_p$ consists of $n$ $Z^\dg$-operators and $m$ $Z$-operators (in some ordering that is not relevant for the moment), i.e. 
\eq[2a]{ W_p = \prod_{i=1}^n Z^\dg_{q_i} \prod_{j=1}^m Z_{r_j} \qquad \text{ with } \qquad \sum_{j=1}^m r_j - \sum_{i=1}^n q_i = p \,\,\,.}
Then we have
\eq{ W_p Z_{p'} \= \(\prod_{i=1}^n S(p',q_i) \prod_{j=1}^m S(r_j,p')\) Z_{p'} W_p \,\,\,,}
where we use the symbol $\=$ to denote equality for the highest order terms only (i.e. we ignore for now all residual lower order terms). To satisfy (\ref{1.1}) we then require that
it holds the equation\footnote{Of course this is not the only way to meet the condition (\ref{1.1}) under (\ref{2a}) but, as the more detailed discussion presented in appendix \ref{app1} shows, the other options lead, at the end, to the same form.}
\eq[3]{ \prod_{i=1}^n S(p',q_i) \prod_{j=1}^m S(r_j,p') = S(p,p')\,\,\,. }
In order to satisfy this relation for all $p,p'$ and also independently of the specific functional form of the $S$-matrix, we have to exploit its general properties. In particular, taking into account that $S(p,p') = S^{-1}(p',p)$ we see that if 
\eq[4.1]{\text{$m=n+1$ and $q_i=r_j$ for all $i = j =1...n$ and $r_{n+1} = p$}}
then eq.\,(\ref{3}) becomes an identity. 

As for the second condition (\ref{2.1}), we have
\eq{ W_p Z^\dg_{p'} \= \(\prod_{i=1}^n S(q_i,p') \prod_{j=1}^m S(p',r_j) \) Z^\dg_{p'} W_p\,\,\,, }
and
\eq{ Z_p W^\dg_{p'} \= \( \prod_{j=1}^m S(r'_j,p) \prod_{i=1}^n S(p,q'_i) \) W^\dg_{p'}Z_{p} \,\,\,,}
and so we would similarly require
\eq{\prod_{i=1}^n S(q_i,p') \prod_{j=1}^m S(p',r_j) = \prod_{j=1}^m S(r'_j,p) \prod_{i=1}^n S(p,q'_i) = S(p',p)\,\,\,.}
Remarkably this condition is essentially identical to the one of eq.\,(\ref{3}), i.e. our solution (\ref{4.1}) of (\ref{3})  automatically satisfies this one too. Hence, there exists a solution to these equations for arbitrarily high order $n+m$.

However this is not the end of the story, since one has also to check the point (b), namely that the residual terms vanish. In order to ensure this condition for all $p,p'$, instead of considering a single product (\ref{2a}), one has to look at a linear combination of such terms for all momenta $q_i$ and choose their coefficients so that the residual terms cancel each other. If this could not be realized, one would still have the option to introduce into $W_p$ suitable lower order terms  and cancel the residual terms order by order. In this way the coefficients of terms of order $n$ depend on those of order $n+2$ and we see that the construction of the transformation can be carried out recursively. 

Provided that all of these requirements are met, the resulting transformation is of the form
\eq[1b]{W_p= \sum_{\{q_i\}} \alpha(p,\{q_i\}) \(\prod_{i=1}^n Z^\dg_{q_i} Z_{q_i}\) Z_p + \text{(suitable lower order terms)}}
Here we only report the lowest order members of this family of transformations. The first one corresponds to $n=1$ and is the one we have found already in the previous section (in this case the residual terms result in a nonzero $T$, as we saw)
\eq{W_p =  i \sum_q a_{p,q} Z^\dg_q Z_q Z_p\,\,\,.}
For $n=2$ we find that the coefficient must be simply an imaginary constant
\eq[ex1b]{W_p = i \sum_{q,r} Z^\dg_q Z_q Z^\dg_r Z_r Z_p \,\,\,.}

Our study started by assuming that $W_p$ is a single product of $Z,Z^\dg$ operators, a monomial (even though later we had to generalize our assumption by considering linear combinations of similar terms and lower order ones). However this is obviously not the only possibility. Another possibility is investigated in appendix \ref{app1}, which starts from a binomial and leads to the discovery of another interesting type of transformations
\ee{W_p & = \sum_q b_q (S_{p,q} S_{p,-q}-1) Z_p Z_q^\dg Z_{-q}^\dg + \sum_q b^*_q (S_{p,q} S_{p,-q}-1) Z_{-q} Z_q Z_p + 2 b_{-p} Z^\dg_{-p} = \nn \\
& = \sum_q b_q (1-S_{q,p} S_{-q,p}) Z_q^\dg Z_{-q}^\dg Z_p + \sum_q b^*_q (S_{p,q} S_{p,-q}-1) Z_{-q} Z_q Z_p - 2 b_{p} Z^\dg_{-p}\,\,\,, \label{ex2}}
where $b_q$ has been chosen to satisfy $b_{-q}=b_q S_{q,-q}$ and we have assumed that the $S$-matrix satisfies $S(0) = S_{p,p} = -1$. As already mentioned this excludes only the case of free bosons since for all other integrable models it is always true. Note that the last term cannot be absorbed by reordering the operators of the first one. 

Let us remark that one may continue in a similar way and construct other more complex classes of transformations. In particular, one may even consider the infinite series of products (\ref{gen_sol}) which, unlike all cases presented above, do not give rise to expressions that terminate at finite order. The study of such transformations will be discussed elsewhere.

\subsection{Properties of the two simple classes of transformations}

In the previous sections we have found mainly two distinct classes of symmetries of the ZF algebra which led to eqs. (\ref{ex1}) and (\ref{ex2}).
The first type of transformations (\ref{ex1}) is the one that generates a change in the $S$-matrix. However this first type does not change the ground state of the theory since $Z'_p=Z_p+\epsilon W_p$ annihilates the same vacuum as $Z_p$. The reason is that this transformation, as well as the second member of the same class (\ref{ex1b}), contain always one more $Z$-operator than $Z^\dg$'s. Finally, let's notice that it does not reduce to the Bogoliubov transformation in the free limit $S=\pm 1$ since it does not depend explicitly on $S$.
\begin{figure}[!ht]
\centering
\includegraphics[width=.4\textwidth]{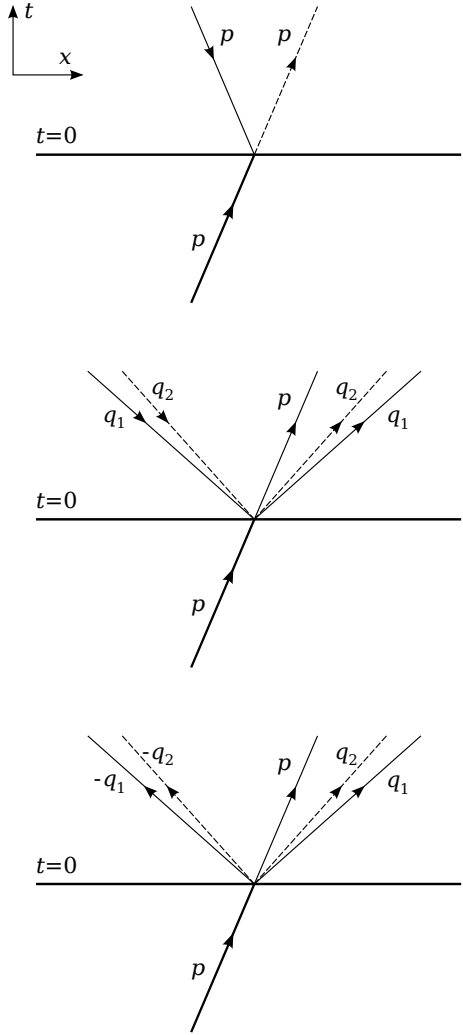}
\caption{\small \emph{Diagrammatic representation of the three types of transformations in the context of quantum quenches. The linear Bogoliubov transformation that works only in free systems corresponds to the transformation of an old particle into a new antiparticle of opposite momentum. The two new types of transformations that work in the interacting case, convert the old particle into a new particle of the same momentum accompanied (or ``dressed'') by a particle-antiparticle (type I) or particle-particle (type II) pair with opposite momenta.}}
\label{fig}
\end{figure}

The second type of transformation (\ref{ex2}) has three important properties. Firstly, it does not change the $S$-matrix. Secondly, in the free limit where $S \to \pm 1$, its nonlinear terms (first and second) vanish, leaving only the linear term (last) $Z^\dg_{-p}$, which corresponds to the Bogoliubov transformation. Thirdly and most importantly, it changes the ground state, since the first and last terms in (\ref{ex2}) contain one more $Z^\dg$ operator than $Z$'s which means that the new annihilation operator does not annihilate the old ground state. In particular, as we will show next, the infinitesimal change in the ground state can be described as creation of a pair of excitations with opposite momenta.

Indeed, if we denote by $|\Omega\ra$ the ground state corresponding to the pre-quench operator $Z$, by definition this state satisfies
\eq{Z_k |\Omega\ra = 0}
for all $k$. This condition, expressed in the basis of the post-quench operator $Z'$ (with corresponding ground state $|\Omega'\ra$), reads
\eq[in_st_cond]{(Z'_k - \ep W_k) (1+ \ep X) |\Omega'\ra = \ep (Z'_k X - W_k) |\Omega'\ra = 0}
where $X$ is a suitable operator to be determined. For $W$ given by (\ref{ex2}) we easily find by normal ordering that
\eq[ssx]{X = - \sum_p b_{p} {Z'}^\dg_{p} {Z'}^\dg_{-p}\,\,\,.}

Notice that this is the infinitesimal version of a squeezed coherent state. In fact we can go much further and show that for real $b_p$ any finite transformation generated by eq.\,(\ref{ex2}) ``transforms'' the initial ground state into a squeezed coherent state\footnote{By ``transformed'' ground state, we mean the expansion of the ground state of the pre-quench operator $Z$ into the basis of the post-quench operator $Z'$.}. To prove this it is sufficient to show that any state $|\Psi\ra$ of the squeezed form
\eq[scs]{|\Psi\ra = \mathcal{N}(K_q) \exp\(\sum_q K_q Z^\dg_q Z^\dg_{-q}\) |\Omega\ra }
in the pre-quench basis preserves its squeezed form under the infinitesimal transformation (\ref{ex2}), i.e. it is ``transformed'' into a squeezed state in the post-quench basis
\eq[cond]{|\Psi\ra = \mathcal{N}(K'_q) \exp\(\sum_q K'_q {Z'}^\dg_q {Z'}^\dg_{-q}\) |\Omega'\ra \,\,\,.}
If this is true then, since the finite transformation can be built up by successive application of infinitesimal ones and the initial ground state is ``transformed'' into a squeezed state (\ref{ssx}) which remains of this form after every infinitesimal step of this procedure, we conclude by induction that the transformation (\ref{ex2}) indeed ``generates'' squeezed states. 

To prove this statement we can follow a path parallel to the corresponding free field calculation. For free bosons or fermions one should show that under an infinitesimal Bogoliubov transformation
\eq[Bog]{Z'_p = Z_p +\epsilon a_p Z^\dg_{-p}, \qquad (a_{-p}=a_p)}
any squeezed state of the form (\ref{scs}) remains squeezed as well. The easiest way to see this is to employ the following equivalent form of (\ref{scs})
\eq[scs2]{|\Psi\ra = \exp\[\sum_q \Lambda_q \(Z^\dg_q Z^\dg_{-q} - Z_{-q} Z_q\) \]|\Omega\ra }
where $\Lambda_q$ is a known function of $K_q$. In the above we have assumed that $a_q, K_q$ and $\Lambda_q$ are all real functions and all of the following are restricted to this case. The operator $(Z^\dg_q Z^\dg_{-q} - Z_{-q} Z_q)$ in the exponent remains invariant under the Bogoliubov transformation (\ref{Bog}) and therefore the only change comes from the ground state
\eq{|\Omega\ra = \(1-\epsilon \sum_q a_q {Z'}^\dg_q {Z'}^\dg_{-q}\)|\Omega\ra'}
which can be absorbed in a shift of the coefficient $\Lambda_q$ in the exponent
\eq{|\Psi\ra = \exp\[\sum_q (\Lambda_q - \epsilon a_q) \({Z'}^\dg_q {Z'}^\dg_{-q} - Z'_{-q} Z'_q\) \]|\Omega'\ra\,\,\,. }
This is exactly what we wished to show. To verify that eq.\,(\ref{scs2}) can also be written in the form (\ref{scs}) one can normal order the squeezing operator $\exp[\sum_q \Lambda_q (Z^\dg_q Z^\dg_{-q} - Z_{-q} Z_q)]$. Alternatively one may first decompose the exponential of the sum over momenta in (\ref{scs2}) into an infinite product of exponentials\footnote{To avoid ordering problems in this product, first restrict the summation variable $q$ to positive values only, as we are allowed to do.}. Then observe that for each pair of opposite momentum modes, the commutation relations of the operators $Z^\dg_q Z^\dg_{-q}, Z_{-q} Z_q$ and $\tfrac{1}{2}(Z^\dg_q Z_q + Z^\dg_{-q} Z_{-q} \pm 1)$ ($+$ for bosons / $-$ for fermions) form a closed algebra ($SU(1,1)$ algebra for bosons / $SU(2)$ algebra for fermions) and therefore we can rewrite the exponential as
\eq{e^{2\Lambda_q (Z^\dg_q Z^\dg_{-q} - Z_{-q} Z_q)} = e^{2K_q(\Lambda_q) Z^\dg_q Z^\dg_{-q}} \; e^{2L_q(\Lambda_q) Z_q Z_{-q}} \; e^{M_q(\Lambda_q) \(Z^\dg_q Z_{q} + Z^\dg_{-q} Z_{-q} \pm 1\)} }
for some appropriate coefficients $K_q,L_q,M_q$ that it is not necessary to determine here. Applying this operator to $|\Omega\ra$, we directly derive the result mentioned above.

Extending this computation to the general integrable case is almost straightforward. The first step is to check that even in this general case, squeezed states can be equivalently written in both forms (\ref{scs}) and (\ref{scs2}). This is true, since the commutation relations of the operators involved in the computation are not crucially different from the free fermionic ones. Indeed, if we define $Y^\dg_q = Z^\dg_q Z^\dg_{-q}$, $N_q = Z^\dg_q Z_q$ and $\bar{N}_q = Z_q Z^\dg_q = - N_q + 1$ then
\ee{
[Y_q^\dg,Y_p^\dg]& = 0 \label{cc1} \\
[Y_q^\dg,Y_p]& = \delta_{q,-p} S_{-p,p} (N_{-p} - \bar{N}_{p}) + \delta_{q,p} (N_{p} - \bar{N}_{-p}) \label{cc2} \\
[Y_q^\dg,N_p]& = - (\delta_{q,-p} + \delta_{q,p}) Y^\dg_{q} \label{cc3}
}
and the only difference with free fermions is the factor $S_{-p,p}$ in (\ref{cc2}). The next step is to check if the operator $(Z^\dg_q Z^\dg_{-q} - Z_{-q} Z_q)$ in the exponent of (\ref{scs2}) remains invariant under the transformation (\ref{ex2}), which turns out, after some algebra, to be true if $b_p$ is real. This completes the proof of our statement.

Let us now discuss another property of this second type of transformations: the change induced to the conserved charges of the theory. As we have already seen in (\ref{cch}), the first type we studied leaves all conserved charges invariant. In the present case the transformation of $Z^\dg_p Z_p$ turns out to be
\eq[cch2]{{Z'}^\dg_{p} {Z'}_{p} = {Z}^\dg_{p} {Z}_{p} - 2 \epsilon (b_p Z_p^\dg Z^\dg_{-p} + b^*_p Z_{-p} Z_p)\,\,\,,}
that is the conserved charges $\hat{\mathcal{Q}}_s$ do change but they remain quadratic in the $Z$ operators. In particular the hamiltonian of the system which in momentum representation and according to our normalization of the ZF operators is $H=\sum_p E_p Z^\dg_p Z_p$ is transformed as
\eq{H' = H + \epsilon \sum_p \[ \frac{\partial E_p}{\partial \epsilon} Z_p^\dg Z_p - 2 E_p (b_p Z_p^\dg Z^\dg_{-p} + b^*_p Z_{-p} Z_p)\]}
where ${\partial E_p}/{\partial \epsilon}$ reflects the change in the dispersion relation induced by the transformation. Notice that this expression is reminiscent of the analogous one for Bogoliubov transformations in free theories.

From a physical point of view and in particular from the perspective of quantum quenches, the action of transformations of these two types consists in ‘dressing’ the initial particle of the theory with a pair of newly created particles with opposite momenta, as depicted in figure \ref{fig}. Especially for the second type, if such a transformation describes a quantum quench, then the initial ground state is expressed as a squeezed coherent state in the post-quench basis, at least for real $b_p$. Finally, the perturbation introduced in the hamiltonian after a small quantum quench of this type would be of quadratic form in the pre-quench ZF operators.

\section{Examples from physical theories}\label{sec_physical}

In this section we will present some first examples of transformations of ZF operators in the context of two important integrable models, the Sinh-Gordon and the Lieb-Liniger. The first is one of the simplest and best-studied relativistic integrable models consisting of a single type of particles, while the second describes a system of non-relativistic interacting bosons in (1+1)d and models experimental cold atom set-ups. As it has been recently shown in \cite{KMT}, the two models are closely related to each other, since the Lieb-Liniger can be obtained as a suitable non-relativistic limit of the Sinh-Gordon model.

In the Sinh-Gordon model, an example of infinitesimal transformation would correspond to a small quench of the mass $m$ or the coupling constant $g$. Starting from arbitrary initial values, such a quench would introduce into the hamiltonian perturbations ($\int dx \; \cosh g\phi(x)$ and $\int dx \; (2 \cosh g\phi(x) +g \phi(x) \sinh g\phi(x))$ respectively) that correspond to an infinite series in terms of the $Z, Z^\dg$ operators, as can be seen from their form factors \cite{KM}. An exception to this rule is when the initial model lies on a free point, $g=0$. In this special case where a small $g$ is abruptly switched on, each term of the expansion is smaller than the previous one by an amount of the order of $g^2$, due to the same property held for the form factors $F^{\phi}_{2n+1}$ of the physical field. Thus the derivation of the corresponding infinitesimal transformation is simpler and can be done by means of perturbation theory, which is what we do in the next section. In the Lieb-Liniger model on the other hand, we use the known relation \cite{Thacker} between the physical field operator and the ZF operators to find the infinitesimal transformation of the latter, again for the case when the interaction changes from zero to a small value. Using the non-relativistic limit mentioned above \cite{KMT} we can verify the consistency of the results for the Sinh-Gordon and Lieb-Liniger models.

Note that in both cases the transformation refers to a small change from a free bosonic to an interacting point of the theory and, according to a comment we made in a footnote of \ref{sub_conditions}, the $S$-matrix at zero momentum $S(0)$ is non-analytic (in fact discontinuous) in the coupling constant at such points. Therefore we anticipate this non-analyticity to become evident in our results and indeed it does as we will see below.

\subsection{The Sinh-Gordon model}

The Sinh-Gordon model is a relativistic field theory in (1+1)d defined by the hamiltonian 
\eq{H = \frac12 \pi^2 + \frac12\left(\frac{\partial\phi}{\partial x}\right)^2 + \frac{m^2 c^2}{g^2}\left(\cosh g\phi-1\right)}
where $\phi=\phi(x,t)$ is a real scalar field, $m$ is a mass scale and $c$ is the speed of light. In this integrable field theory there is only one type of particle with physical (renormalized) mass $M$ given by
\eq[mass]{M^2 = m^2 \frac{\sin \alpha \pi}{\alpha \pi} }
where $\alpha$ is the dimensionless renormalized coupling constant
\eq{\alpha = \frac{ c g^2}{8\pi+ c g^2}}
Particle scattering is fully determined by the two-particle $S$-matrix given by
\begin{equation}
S_{\text{sh-G}}(\theta,\alpha)=\frac{\sinh\theta-i\,\sin \alpha\pi}{\sinh\theta+i\,\sin \alpha\pi} \label{eq_sinh_S}
\end{equation}
where $\theta$ is the rapidity difference between the particles.

To calculate the ZF operators from first order perturbation theory we consider the $\phi^4$ model 
\eq{H = \frac12 \pi^2 + \frac12\left(\frac{\partial\phi}{\partial x}\right)^2 + \frac12 {m^2 c^2} \phi^2 + \frac{\lambda}{4!} \phi^4}
with coupling constant $\lambda=m^2 c^2 g^2$ (we set $c=1$ from now on). We first define the auxilliary operators 
\eq[BOp]{B^\dg_+(k) = \Omega_+ A^\dg(k) \Omega_+^\dg \quad , \qquad B^\dg_-(k) = \Omega_- A^\dg(k) \Omega_-^\dg}
where $\Omega_\pm$ are the following evolution operators (M{\o}ller operators)
\eq{\Omega_\pm = \lim_{T\to\pm\infty}e^{-i\int^0_{-T}{dt H_{int}(t)}}}
As known from the general scattering theory, the operators $ B^\dg_+(k) , B^\dg_-(k)$ when acting on the vacuum state of the interacting theory $|\Omega\ra$, create ``in'' and ``out'' states respectively. If we consider the interaction hamiltonian $H_{int}$ as normal-ordered, we have
\ee{B^\dg_\pm(k) & = A^\dg (k) - i \frac{\lambda}{4!} \int^0_{\mp \infty} dt \int dx \[:\phi^4(x,t):,A^\dg(k)\] = \nn \\
& = A^\dg(k) - i \frac{\lambda}{3!} \int^0_{\mp \infty} dt  \int dx \frac{e^{-iE_k t+ikx}}{\sqrt{2E_k}} :\phi^3(x,t): \label{eqB2}}
and expanding in terms of the free boson creation/annihilation operators $A(k), A^\dg(k)$
\ee{& B^\dg_\pm(k) = A^\dg(k) - \nn \\ 
& - i \frac{\lambda}{3!} \int^0_{\mp \infty} dt \frac{e^{-iE_k t}}{\sqrt{2E_k}} \int \frac{dk_1 dk_2 dk_3  2\pi\delta(k+{\textstyle \sum_i k_i})}{(2\pi)^3 \sqrt{2^3 E_{k_1} E_{k_2} E_{k_3}}} \Bigg[ A(k_1) A(k_2) A(k_3) e^{-i(E_{k_1}+E_{k_2}+E_{k_3})t} + \nn \\
& + 3 A^\dg(-k_1) A(k_2) A(k_3) e^{-i(-E_{k_1}+E_{k_2}+E_{k_3})t} + 3 A^\dg(-k_1) A^\dg(-k_2) A(k_3) e^{-i(-E_{k_1}-E_{k_2}+E_{k_3})t} + \nn \\
& + A^\dg(-k_1) A^\dg(-k_2) A^\dg(-k_3) e^{-i(-E_{k_1}-E_{k_2}-E_{k_3})t} \Bigg] }
In scattering theory the $t$-integration is understood under the ``adiabatic switching'' prescription which means to introduce an $ e^{-\epsilon |t|} $ factor into the integrand. According to the identity
\eq{\int^0_{\pm\infty} dt \; e^{-\epsilon |t| - i \omega t} = \frac{i}{\omega \mp i \epsilon}}
we then find
\ee{ & B^\dg_\pm(k) = A^\dg(k) + \frac{\lambda}{3!} \frac{1}{\sqrt{2E_k}} \int \frac{dk_1 dk_2 dk_3  2\pi\delta(k+{\textstyle \sum_i k_i})}{(2\pi)^3 \sqrt{2^3 E_{k_1} E_{k_2} E_{k_3}}} \Bigg[ \frac{A(k_1) A(k_2) A(k_3)}{E_k + E_{k_1}+E_{k_2}+E_{k_3}} + \nn \\
& + 3 \frac{A^\dg(-k_1) A(k_2) A(k_3)}{E_k -E_{k_1}+E_{k_2}+E_{k_3}} + 3 \frac{A^\dg(-k_1) A^\dg(-k_2) A(k_3)}{E_k -E_{k_1}-E_{k_2}+E_{k_3} \pm i\epsilon} + \frac{A^\dg(-k_1) A^\dg(-k_2) A^\dg(-k_3)}{E_k -E_{k_1}-E_{k_2}-E_{k_3}} \Bigg] }
Notice that we kept the $\pm i\epsilon$ shift only in the 3rd term since this is the only one that has a singularity (at $k_1=-k,k_2=-k_3$ or $k_2=-k,k_1=-k_3$). Using the formal identity
\eq[id2]{\lim_{\epsilon\to 0^+}\frac{1}{\omega \pm i\epsilon}=\mathcal{P}
\left(\frac{1}{\omega}\right) \mp i\pi \, \delta(\omega)}
we can rewrite the above as
\eq{B_\pm^\dg(k) = B^\dg(k) \mp \frac{i \lambda}{8} \int \frac{dq}{2\pi} \frac{1}{|k E_q - q E_k|} A^\dg(k) A^\dg(q) A(q) }
where 
\ee{ & B^\dg(k) \equiv A^\dg(k) + \frac{\lambda}{3!} \frac{1}{\sqrt{2E_k}} \int \frac{dk_1 dk_2 dk_3  2\pi\delta(k+{\textstyle \sum_i k_i})}{(2\pi)^3 \sqrt{2^3 E_{k_1} E_{k_2} E_{k_3}}} \Bigg[ \frac{A(k_1) A(k_2) A(k_3)}{E_k + E_{k_1}+E_{k_2}+E_{k_3}} + \nn \\
& + 3 \frac{A^\dg(-k_1) A(k_2) A(k_3)}{E_k -E_{k_1}+E_{k_2}+E_{k_3}} + 3 \; \mathcal{P.V.} \frac{A^\dg(-k_1) A^\dg(-k_2) A(k_3)}{E_k -E_{k_1}-E_{k_2}+E_{k_3}} + \frac{A^\dg(-k_1) A^\dg(-k_2) A^\dg(-k_3)}{E_k -E_{k_1}-E_{k_2}-E_{k_3}} \Bigg]}

From their definition (\ref{BOp}), each of the two operators $B_\pm^\dg(k)$ satisfy the same standard commutation relations as the free operators $A(k),A^\dg(k)$. Now let us define the operator
\ee{ Z^\dg(k) & \equiv B^\dg(k) - \frac{i \lambda}{8} \int \frac{dq}{2\pi} \(\frac{1}{k E_q - q E_k}\) A^\dg(k) A^\dg(q) A(q) = \nn \\ 
& = B_\pm^\dg(k) - \frac{i \lambda}{8} \int \frac{dq}{2\pi} \(\frac{1}{k E_q - q E_k} \mp \frac{1}{|k E_q - q E_k|} \) A^\dg(k) A^\dg(q) A(q) \label{ZFshG}}
Since this is of the form (\ref{ex1}), we automatically know that $Z^\dg(k)$ and $Z(k)$ satisfy the ZF algebra with non-trivial $S$-matrix given by
\eq{S(k,q) = 1 - \frac{i \lambda}{4 (k E_q- q E_k)}}
which is indeed the correct first order perturbation to the $S$-matrix (\ref{eq_sinh_S})
\eq[Smat]{S_{\text{sh-G}}(\theta,\alpha) = 1 - c g^2 \frac{i}{4\sinh \theta}+ \mathcal{O} (c^2 g^4)} 
Notice the infrared singularity in the coefficients of (\ref{ZFshG}) and (\ref{Smat}) when the momentum difference $k-q$ tends to zero. This reflects the non-analyticity of $S(0)$ as $g\to 0$ for which we talked in the introduction of this section. 

Next we consider the states created by the action of $Z^\dg(k)$ on the perturbed vacuum $|\Omega\ra$ which by a calculation similar to the ones above, turns out to be
\eq{|\Omega\ra = \Omega_\pm |0\ra = \(1 - \frac{\lambda}{4!} \int {\textstyle \prod_i^4 dk_i} \; \frac{2\pi\delta({\textstyle \sum_i^4 k_i})}{(2\pi)^4 \sqrt{2^4 {\textstyle \prod_i^4 E_{k_i}}}} \frac{A^\dg(k_1) A^\dg(k_2) A^\dg(k_3) A^\dg(k_4)}{{\textstyle \sum_i^4 E_{k_i}}} \) |0\ra }
For the one-particle states it can be immediately seen that $Z^\dg(k)|\Omega\ra = B_{\pm}^\dg(k)|\Omega\ra $ always up to first order in $\lambda$. However, in order to verify that $Z^\dg(k)$ plays the right role in creating in and out scattering states, we should check the two-particle states $Z^\dg(k_1)Z^\dg(k_2)|\Omega\ra$. By normal ordering we find
\eq{Z^\dg(k_1) Z^\dg(k_2) |\Omega\ra = B_\pm^\dg(k_1) B_\pm^\dg(k_2) |\Omega\ra - \frac{i \lambda}{8} \(\frac{1}{k_1 E_{k_2} - k_2 E_{k_1}} \mp \frac{1}{|k_1 E_{k_2} - k_2 E_{k_1}|} \) A^\dg(k_1) A^\dg(k_2) |0\ra }
Observing that $k/E_k$ is a monotonically increasing function of $k$, we can easily see that, if $k_1>k_2$ then
\eq{Z^\dg(k_1) Z^\dg(k_2) |\Omega\ra = B_+^\dg(k_1) B_+^\dg(k_2) |\Omega\ra}
i.e. it defines an in state, while if $k_1<k_2$ then
\eq{Z^\dg(k_1) Z^\dg(k_2) |\Omega\ra = B_-^\dg(k_1) B_-^\dg(k_2) |\Omega\ra}
i.e. it defines an out state.

This example, apart from demonstrating how ZF operators emerge from standard perturbation theory of scattering, it also illustrates the concepts developed before and in particular the role of the first class of transformations (\ref{ex1}) that we derived abstractly. We close our presentation of physical examples with the Lieb-Liniger model. We will also verify the consistency of our results for the two models, under the double non-relativistic limit that reduces the former to the latter.

\subsection{The Lieb-Liniger model}

The Lieb-Liniger model describes a (1+1)d system of non-relativistic bosons interacting with each other with a $\delta$-function potential. Its hamiltonian in second quantized form is
\eq{H = \int_{-L}^{+L} dx \; \( \frac{1}{2} \partial_x\Psi^\dagger(x) \partial_x\Psi(x) + \lambda \; \Psi^\dagger(x) \Psi^\dagger(x) \Psi(x) \Psi(x) \)\,\,\,,}
where $\lambda$ now is the interaction strength. The ground state energy for a system of $N$ bosons as well as its thermodynamics can be exactly worked out by means of Bethe ansatz \cite{LL}. The exact solution expresses the energy of the ground state and the excitation spectrum in terms of the dimensionless coupling constant $\gamma \equiv \lambda/\rho$ where $\rho = N/L$ is the density of bosons with $N,L\to\infty$.

In this model the relation between the bosonic field operator $\Psi(x)$ and the ZF operators $R_{\lambda}(k)$ that diagonalize the hamiltonian for $L\to\infty$ has been already found using the inverse scattering method \cite{Thacker}
\eq[Thacker_tr]{\Psi(x)=\sum_{N=0}^{\infty}\int\prod_{i=1}^{N}\frac{dp_{i}}{2\pi}\prod_{j=0}^{N}\frac{dk_{j}}{2\pi} \; g_{N}(\{p\},\{k\};x) \; R_{\lambda}^{\dg}(p_{1})\cdots
R_{\lambda}^{\dg}(p_{N})R_{\lambda}(k_{N})\cdots R_{\lambda}(k_{1}) R_{\lambda}(k_{0})}
where
\eq{g_{N}(\{p\},\{k\};x)=\frac{(-\lambda)^{N}\exp[i(\sum_{i=0}^{N}k_{i}-\sum_{i=1}^{N}p_{i})x]}{\prod_{j=1}^{N}(p_j-k_j-i\epsilon)(p_j-k_{j-1}-i\epsilon)}\,\,\,.}
Indeed it can be shown that the $R,R^\dg$ operators diagonalize the hamiltonian and satisfy the ZF algebra
\begin{eqnarray}
[H,R_{\lambda}^{\dg}(q)] &=& q^{2}R_{\lambda}^{\dg}(q),\\ 
 R_{\lambda}(q)R_{\lambda}(q') &=& S_{\lambda}(q'-q)R_{\lambda}(q')R_{\lambda}(q), \\ 
R_{\lambda}(q)R_{\lambda}^{\dg}(q')& = & S_{\lambda}(q-q')R_{\lambda}^{\dg}(q')R_{\lambda}(q)+2\pi \delta(q-q')
\end{eqnarray}
where the $S$-matrix is
\eq{S_\lambda(q)=\frac{q -i \lambda}{q + i \lambda}\,\,\,.}

Let us consider the infinitesimal transformation from the free bosonic point $\lambda=0$ to a small value $\lambda$. From (\ref{Thacker_tr}) we have\footnote{Note that, unlike in a relativistic free field theory where the creation-annihilation operators are linear combinations of the field $\phi$ and its conjugate momentum $\pi$, in a non-relativistic free field theory the creation-annihilation operators are the bosonic field itself $R_{0}(k)=\int dx \; e^{-ikx} \Psi(x)$. Also the conjugate momentum is $\Pi=i\Psi^\dg$ and does not appear in the hamiltonian.}
\eq[tr1]{R_{\lambda}(k)= R_0(k) + \lambda \int \frac{dq dq'}{(2\pi)^2} \; \frac{R_0^{\dg}(q+q'-k)R_0(q)R_0(q')}{(q-k-i\epsilon)(q'-k-i\epsilon)} + \mathcal{O}(\lambda^2)}
The $S$-matrix is no longer unit but becomes instead
\eq{S_{\lambda}(p) = 1 - \lambda\frac{2i}{p} + \mathcal{O}(\lambda^2) \,\,\,}
and so, according to our previous findings, we expect that the transformation contains the generator of $S$-matrix shifts (\ref{ex1}) with coefficients $a_{k,q}=i/(q-k)$. Indeed, using the identity (\ref{id2}) we recognize that part of the infinitesimal transformation (\ref{tr1}) has exactly the form of (\ref{ex1}) with the right coefficient
\eq{ \int \frac{dq}{2\pi} \; \frac{i}{q-k} R_0^{\dg}(q)R_0(q) R_0(k)\,\,\,,}
while the remaining part does not affect the $S$-matrix. Once again, notice the infrared singularity in the coefficient of the above expression for $q=k$.

Lastly, we mention that the infinitesimal transformations (\ref{ZFshG}) and (\ref{tr1}) derived for the Sinh-Gordon and Lieb-Liniger model respectively, are consistent with each other under the double non-relativistic limit $c\to \infty$, $g\to 0$, $g c: const.$, that leads from the former to the latter model. Following \cite{KMT} we substitute the field $\phi$ in (\ref{eqB2}) as
\eq{\phi(x,t)=\frac{1}{\sqrt{2m}}\left(\psi(x,t) e^{-i m c^2 t}+\psi^\dg(x,t) e^{+i m c^2 t}\right)}
and keep only the non-oscillating terms, rewriting first all expressions with their $c$ dependence explicit and taking into account that in the non-relativistic limit $E_k = m c^2 + k^2/2m +...$. After some algebra we verify that (\ref{ZFshG}) reduces to (\ref{tr1}).

\section{Conclusions}

In this article we have investigated how the initial state of a quantum quench in an integrable model can be expressed, from first principles, in terms of ZF operators, without relying on the usual mapping to slab geometry and the associated boundary renormalization group arguments \cite{CC1,CC2,FM}. We show that this result can be achieved by deriving the relation between the pre-quench and the post-quench operators on the condition that such a relation respects the Zamolodchikov-Faddeev (ZF) algebra satisfied by integrable models.

Under the conditions that such transformations must satisfy the ZF algebra at the infinitesimal order, we have initially showed that the usual linear Bogoliubov transformations do not respect the ZF algebra, apart from the trivial cases of free bosons or fermions, a result that holds generally for finite transformations too. 

We have then identified two important classes of transformations. The first class changes the $S$-matrix of the theory but preserves its ground state as well as its conserved charges. We also argued that any infinitesimal transformation can be decomposed into a part that induces the $S$-matrix shift and the rest that does not alter the $S$-matrix. The second class belongs to the latter subset of transformations, which can be regarded as a generalization of the Bogoliubov transformations for interacting theories, since it reduces to the usual Bogoliubov transformations whenever the integrable model reaches a free bosonic or fermionic point. Like in the free case, the ground state of the system becomes a squeezed state when expressed in the transformed ZF basis under such a generalized Bogoliubov transformation, at least when its coefficients are real. We have also showed that the change in the hamiltonian (and in the other conserved charges of the theory) is of the same form as that of the Bogoliubov transformations. The net effect of this type of transformations is to ``dress'' the initial quasiparticle with pairs of new particles with momenta opposite to each other. 

We have also outlined how one could proceed further in this program to identify the transformations which preserve the ZF algebra, in particular pointing out the existence of transformations of higher complexity characterized by the fact that, even for infinitesimal quenches, they are associated to an \emph{infinite} series of terms given by products of the initial creation/annihilation operators. In the quantum quench perspective, this means that even a small quench of the physical parameters of an integrable model may result in an infinite series which links the pre-quench and the post-quench operators. In this case, the calculation of the initial state made on first principles is rather difficult, unless a truncation or resummation of the series can be established on the grounds of a different argument. In such a case, for instance, it may be possible to reorganize the terms of the series based on a small-density expansion, following the concepts developed in \cite{Essler}. We hope that our work on ZF algebra transformations will stimulate further investigation of their structure and properties, both from a pure and an applied point of view.

Lastly we exemplified our approach in the context of the Sinh-Gordon and Lieb-Liniger model. We restricted ourselves to perturbations about the free bosonic point of these models since in this case the transformations can be found relatively easily and contain only up to cubic terms in the ZF operators. We expect that analogous simplification occurs near free points of other integrable models too and it would be interesting to explore some physical realization of such quench processes. Regarding the Sinh-Gordon model, an application of our results to the corresponding quantum quench problem of an abrupt switch-on of the interaction, would give results comparable with earlier work \cite{Sotiriadis}. For the Lieb-Liniger model instead, further manipulation is required, mainly due to the fact that the ground state is not the empty vacuum but contains a large number of particles proportional to the size of the system. This issue is discussed and a numerical approximation is developed in \cite{Gritsev}. A recent numerical study of a special quantum quench in the Lieb-Liniger model is \cite{MC}.

\vspace{15mm} {\em Acknowledgements}. We wish to thank G. Delfino, F. Essler and A. Silva for helpful discussions.

\newpage

\appendix
\section{Squeezed states in free quantum field theories}

In this Appendix we show that the squeezed coherent states in a quantum quench of free theories are a consequence of the Bogoliubov transformation of their operators. 

{\bf Bosonic theory.} Let us consider firstly the quench in a bosonic theory with hamiltonian
\cite{CC2} 
$$
H\,=\,\frac{1}{2} \int \left[\pi^2 + (\nabla \varphi)^2 + m_0^2 \varphi^2\right] \, dx \,\,\,.
$$
This system can be diagonalized in momentum space  
\begin{eqnarray*}
&& H\,=\,\int \Omega_k^0 \,:A^{0 \dagger}_k A^0_k: \,\,\,, \\
&&(\Omega^0_k)^2\,=\,m_0^2+k^2,\\
&& A^0_k=\frac {1}{\sqrt{2 \Omega^0_k}} \left(\Omega^0_k \varphi_k+i \pi_k \right)\,\,\,,
\\
&& A^{0 \dagger}_k=\frac {1}{\sqrt{2 \Omega^0_k}} \left(\Omega^0_k \varphi_{-k}-i \pi_{-k} \right)\,\,\,, 
\end{eqnarray*}
with the ground state $|\Psi_0\rangle$ identified by the condition 
\EQ
A^0_k |\Psi_0\rangle \,=\,0 \,\,\,.
\label{groundstateB}
\EN
Imagine now that, after having prepared the system in its ground state, we quench 
the mass $m_0 \rightarrow m$.  The relation between the pre-quench ladder operators $(A^0_k,A^{0 \dagger} _k)$  and the post-quench ones $(A_k,A^{\dagger}_k)$ is a Bogoliubov trasformation
\begin{eqnarray*}
&& A_k\,=\,c_kA^0_k+d_k A^{0 \dagger}_{-k},\qquad A^\dagger_k=c_kA^{0 \dagger }_k + d_k A^{0}_{-k} \\
&& A^0_k\,=\,c_kA_k-d_k A^{ \dagger}_{-k},\qquad A^{0\dagger}_k=c_kA^{\dagger }_k - d_k A_{-k}, \label{bosonic_bogoliubov}
\end{eqnarray*}
where the coefficients are given by 
$$
c_k=\frac { \Omega_k+\Omega_k^0} {2 \sqrt{\Omega_k\Omega^0_k}}, \qquad d_k=\frac { \Omega_k-\Omega_k^0} {2 \sqrt{\Omega_k\Omega^0_k}}.
$$
Substituting the expression of $A^0_k$ from the Bogoliubov transformation into eq.\,(\ref{groundstateB}), we see that, in terms of the new operators, the initial state satisfies the condition 
\EQ
\left[c_kA_k-d_k A^{ \dagger}_{-k}\right]|\Psi_0\rangle \,=\,0 \,\,\,.
\label{groundstateB1}
\EN
whose solution is given in terms of a squeezed coherent state
\begin{equation}
|\Psi_0\rangle=\mathcal{N} \exp \left[\int_{-\infty}^{\infty} K_{boson}(k) A_k^\dg A_{-k}^\dg dk\right] |0\rangle , \label{bosonic_state}
\end{equation}
where
\begin{equation} 
 K_{boson}(k) \,=\, \frac{\Omega^0_k-\Omega_{k}}{\Omega^0_k+\Omega_{k}}
\label{Kk}
\end{equation}
This quantity can be written in a suggestive way by introducing the rapidities of the particle relative to the initial and final situations, i.e. 
\begin{eqnarray}
&\Omega^0 = m_0 \cosh\theta_0 
\,\,\,\,\,\,
,
\,\,\,\,\,\,
& k = m_0 \sinh\theta_0 \label{rapidities} \\
&\Omega = m \cosh\theta 
\,\,\,\,\,\,
,
\,\,\,\,\,\,
& k = m \sinh\theta \nonumber 
\end{eqnarray}
From the equality of the initial and final momenta, we have the relation which links the 
two rapidities
\begin{equation}
m_0 \sinh\theta_0 = m \sinh\theta \,\,\,\,\, 
\Rightarrow  
\frac{m_0}{m} = \frac{\sinh\theta}{\sinh\theta_0}
\end{equation}
and therefore, the amplitude $K_{boson}(k)$ of eq.\,(\ref{Kk}) can be neatly written as  
\begin{eqnarray}
K_{boson}(\theta,\theta_0) & = & \frac{m_0 \cosh\theta_0 - m \cosh\theta}{m_0 \cosh\theta_0 + m \cosh\theta} \,=\,
\frac{\frac{m_0}{m}\cosh\theta_0 -  \cosh\theta}{\frac{m_0}{m}\cosh\theta_0 +  \cosh\theta} = \\
& = & \frac{\sinh\theta \cosh\theta_0 - \sinh\theta_0 \cosh\theta}{\sinh\theta \cosh\theta_0 + \sinh\theta_0 \cosh\theta} 
= \frac{\sinh(\theta - \theta_0)}{\sinh(\theta + \theta_0)} \,\,\,. \nonumber 
\end{eqnarray}
 
 \vspace{1mm}
{\bf Fermionic theory}. 
One can easily work out the Bogoliubov transformation relative to the quench of the mass of a free fermionic system \cite{Rossini}. Consider, in particular, 
a free Majorana fermion in (1+1) dimensions, with the mode expansion of the two components of this field given by 
\begin{eqnarray*}
\psi_1(x,t) & = & \int_{-\infty}^{+\infty} dp \left[\alpha(p) A(p) e^{-i E t + i p x} + 
\overline\alpha(p) A^{\dagger}(p) e^{i E t - i p x} \right] 
\label{fermionM}\\
\psi_2(x,t) & = & \int_{-\infty}^{+\infty} dp \left[\beta(p) A(p) e^{-i E t + i p x} + 
\overline\beta(p) A^{\dagger}(p) e^{i E t - i p x} \right] \nonumber
\end{eqnarray*}
where 
\begin{eqnarray*}
&&\alpha(p) = \frac{\omega}{2\pi \sqrt{2}} \frac{\sqrt{E+p}}{E} 
\,\,\,\,\,\,\,\,\,
, 
\,\,\,\,\,\,\,\,\,
\overline{\alpha}(p) = \frac{\overline{\omega}}{2\pi \sqrt{2}} \frac{\sqrt{E+p}}{E}  
\label{alphabeta} \\
&&\beta(p) = \frac{\overline{\omega}}{2\pi \sqrt{2}} \frac{\sqrt{E-p}}{E} 
\,\,\,\,\,\,\,\,\,
, 
\,\,\,\,\,\,\,\,\,
 \overline\beta(p) = \frac{\omega}{2\pi \sqrt{2}} \frac{\sqrt{E-p}}{E} \nonumber 
\end{eqnarray*}
with $\omega = \exp(i \pi/4)$. At $t=0$, i.e. at the instant of the quench, we can extract the 
Fourier mode of each component of the Majorana field 
$
\psi_i(x,0) \,=\,\int dp \,\hat\psi_i(p) e^{i p x} \,\,\,,
$
given by  
\begin{eqnarray*}
\hat\psi_1(p) & = & \alpha(p) A(p) + \overline\alpha(-p) A^{\dagger}(-p) 
\\
\hat\psi_2(p) & = & \beta(p) A(p) + \overline\beta(-p) A^{\dagger}(-p)  
\nonumber 
\end{eqnarray*}
Suppose now that the mass of the field is changed from $m_0$ to $m$ at $t=0$ and 
let's denote by $(A_0(p),A_0^{\dagger}(p))$ and $(A(p),A^{\dagger}(p))$ the sets of oscillators 
before and after the quench. The proper boundary condition associated to such a situation is the continuity of the field components before and after the quench, i.e.   
$
\psi_i^0(x,t=0) \,=\,\psi_i(x,t=0)
$,
which implies 
$
\hat\psi_i^0(p) \,=\,\hat\psi_i(p)
$. 
This gives rise to the Bogoliubov transformation between the two sets of oscillators
\begin{eqnarray*}
&& A_0(p) = u(p) A(p) + i v(p) A^{\dagger}(-p) 
\label{bogfer}\\
&& A_0^{\dagger}(p) = u(p) A^{\dagger}(p) - i v(p) A(-p) \nonumber 
\end{eqnarray*}
where 
\begin{eqnarray*}
&& u(p) \,=\,\frac{1}{2 E} \left[\sqrt{(E_0 +p) (E + p)} + \sqrt{(E_0 - p)(E - p)}\right] 
\label{u&v}\\
&& v(p) \,=\,\frac{1}{2 E} \left[\sqrt{(E_0 - p) (E + p)} - \sqrt{(E_0 + p)(E - p)}\right] 
\nonumber 
\end{eqnarray*}
Notice that these functions satisfy the relations $u(p) = u(-p)$ and $v(p) = - v(-p)$ together with $u^2(p) + v^2(p) = E_0/E$, which refers to the normalization of the respective set of 
oscillators. 

With the same procedure used in the bosonic case, it is easy to see that 
the boundary state corresponding to this quench can be written as 
$$
| B \ra = \exp\(\int_{-\infty}^{\infty} dp \; K_{fermion}(p) A^\dg(p) A^\dg(-p)\) | 0 \ra\,\,\,,
\label{squeezedferm}
$$
where 
\begin{equation}
K_{fermion}(p) \,=\, - K_{fermion}(-p)\,=\, i \frac{\sqrt{(E_0 - p) (E +p)} - \sqrt{(E_0 + p) (E - p)}}
{\sqrt{(E_0 + p) (E +p)} + \sqrt{(E_0 - p) (E - p)}}
\label{kfermion}
\end{equation}
As in the bosonic case, this quantity can be expressed in a more concise form by introducing the rapidities of the particle before and after the quench, i.e.
$$
E_0 \pm p = m_0 e^{\pm \theta_0} 
\,\,\,\,\,\,\,
,
\,\,\,\,\,\,\,
E \pm p = m e^{\pm \theta}
\,\,\,.
$$
Substituting these expressions in (\ref{kfermion}), we get 
\begin{equation}
K_{fermion}(\theta,\theta_0) \,=\,i\,\frac{\sinh\left(\frac{\theta-\theta_0}{2}\right)}
{\cosh\left(\frac{\theta+\theta_0}{2}\right)} \,\,\,. 
\end{equation}
In conclusion, the squeezed coherent form of the initial state in free theories comes from the fact that, in any quantum quench of these systems, the creation-annihilation operators before and after it, are related by a Bogoliubov transformation. And this in turn is a consequence of the canonical commutation/anticommutation relations satisfied by these fields\footnote{To be precise there can be exceptions to this rule since it is possible to construct generalized Bogoliubov transformations which satisfy the CCR/CAR but are nonlinear \cite{nonlinBog1,nonlinBog2,nonlinBog3}. These however correspond to non-quadratic hamiltonians which, even though they can be reduced to free ones, they are uncommon in physically interesting cases.}. 

\begin{figure}[tbh!]
\centering
\includegraphics[totalheight=0.2\textheight]{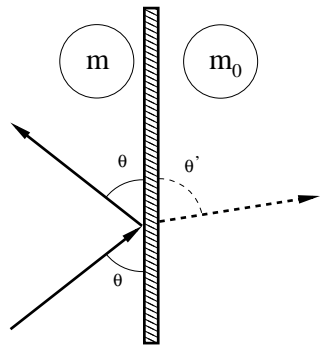} \label{fig_mass_quench}
\caption{{\em The boundary conditon for a mass quench allows for transmission. Notice that, for free theories, there is no particle production at the boundary and the transmitted particle has always the same momentum as the incoming one, since momentum is conserved at the boundary. However, since the mass is different at $x<0$ and $x>0$, the rapidity changes from $\theta$ to $\theta'$.}}
\end{figure}

Finally, notice that both the bosonic and fermionic amplitudes $K(p)$ do not satisfy, in general, a unitarity equation, in contrast to the amplitudes of squeezed coherent states in the purely boundary integrable theories studied by Ghoshal and Zamolodchikov \cite{Ghoshal}. 
The reason why this condition is requested for purely boundary field theory but not for an arbitrary quench process is quite easy to understand. In purely boundary theory, all the degrees of freedom for $t < 0$ are completely frozen: the hard-wall boundary does not allow any transmission process through the boundary and this ends up in the unitarity condition. However, for quantum quenches in free theories, there are degrees of freedom also for $t < 0$, which are related to the ones at $t > 0$ just by the Bogoliubov transformations. Said differently, the boundary condition in free theories allows for transmission, as it is shown in figure (\ref{fig_mass_quench}). The degrees of freedom present on both sides of the boundary prevent the bosonic and fermionic amplitudes (\ref{Kk}) and (\ref{kfermion}) to satisfy a unitarity equation. Notice that the only case where they satisfy a unitarity equation is when we freeze the degrees of freedom before the quench by taking the limit $m_0 \rightarrow \infty$: in this limit, in fact, we have a transmission-less boundary which implements in both theories the Dirichelet boundary condition, with $K_{boson}=1$ and 
$K_{fermion} = i \tanh\frac{\theta}{2}$ respectively. 

\section{Derivation of two classes of generators of ZF algebra transformations}\label{app1}

In this appendix we present the derivation of the two simple classes of infinitesimal transformations of the ZF algebra that we introduced in the main text. As already mentioned we can always write $W$ as a linear combination of products of $Z$ and $Z^\dg$. The first class arises when we consider $W$ to be a single such product
\eq[appW]{W_p=\prod Z_i \prod Z_j^\dg \qquad \text{with} \qquad p=\sum p_i - \sum p_j}
For brevity we use the notation $Z_i \equiv Z_{p_i}$ when there is no confusion about the meaning of the indices, and also the symbol `$\=$' that we defined in paragraph \ref{sec1}. Then we find
\eq{W_1 Z_2 \= \lambda_{12} Z_2 W_1}
where $\lambda_{12} = \prod S_{i2} \prod S_{2j}$ (note the implicit dependence on $p_1$ through the momentum condition in (\ref{appW})). Similarly we have
\eq{W_1 Z_2^\dg \= \lambda_{12}^* Z_2^\dg W_1}
From (\ref{1}) and (\ref{2}) we have the conditions 
\begin{align}
& W_1 Z_2 + Z_1 W_2 \=
S_{12}(Z_2 W_1 + W_2 Z_1) \label{b1} \\
& W_1 Z^\dg_2 + Z_1 W^\dg_2 \= 
S_{21}(Z^\dg_2 W_1 + W^\dg_2 Z_1) \label{b2}
\end{align}
where we ignored $T_{12}$ since the corresponding terms are not of highest order (unless $W$ is a single operator instead of a product, which is the linear case that has already been excluded as not fulfilling the conditions). In the main text eq.(\ref{3}), we considered the simple choice $\lambda_{12}=S_{12}$ which automatically satisfies the above condition but may not be the only possibility. Indeed (\ref{b1}) is in general a weaker condition than this choice. Strictly speaking the two conditions above require
\begin{align}
& (\l_{12}-S_{12}) Z_2 W_1 + (\l^*_{21}-S_{12}) W_2 Z_1 \= 0 \\
& (\l^*_{12}-S_{21}) Z_2^\dg W_1 + (\l_{21}-S_{21}) W_2^\dg Z_1 \= 0
\end{align}
which means, either $\lambda_{12}=S_{12}$ as before, or that $W_2 Z_1$ and $Z_2 W_1$ contain exactly the same operators and similarly for $W_2^\dg Z_1$ and $Z_2^\dg W_1$. In this last case $W_p$ must contain $Z_p$ i.e. $W_p = Z_p \Wt$ where $\Wt$ has momentum zero and, as it can be easily deduced from the above equations, it must be independent of $p$ and contain the same operators as its hermitian conjugate $\Wt^\dg$, that is $\Wt\=\prod Z_i^\dg Z_i$. But this is again exactly the same case considered before (\ref{4.1}) and corresponding to $\lambda_{12}=S_{12}$.

Another possibility arises when we consider $W_p$ to be a linear combination of two products of operators of the previous form
\eq{W_p=a_{+,p} W_{+,p}+ a_{-,p} W_{-,p}}
The conditions are then
\begin{align}
& a_{+1}W_{+1} Z_2 + a_{+2}Z_1 W_{+2} + a_{-1}W_{-1} Z_2 + a_{-2}Z_1 W_{-2} \= \nn \\ 
& \quad S_{12}(a_{+1}Z_2 W_{+1} + a_{+2}W_{+2} Z_1 + a_{-1}Z_2 W_{-1} + a_{-2}W_{-2} Z_1) \label{c1} \\
& a_{+1}W_{+1} Z_2^\dg + a^*_{+2}Z_1 W_{+2}^\dg + a_{-1}W_{-1} Z_2^\dg + a^*_{-2}Z_1 W_{-2}^\dg \= \nn \\ 
& \quad S_{21}(a_{+1}Z_2^\dg W_{+1} + a^*_{+2}W_{+2}^\dg Z_1 + a_{-1}Z_2^\dg W_{-1} + a^*_{-2}W_{-2}^\dg Z_1) \label{c2}
\end{align}
or
\begin{align}
& (\l_{+12}-S_{12})a_{+1}Z_2 W_{+1} + (\l^*_{+21}-S_{12})a_{+2}W_{+2} Z_1 + \nn \\ 
& \qquad (\l_{-12}-S_{12})a_{-1}Z_2 W_{-1} + (\l^*_{-21}-S_{12})a_{-2}W_{-2} Z_1 \= 0 \label{d1} \\
& (\l^*_{+12}-S_{21})a_{+1}Z_2^\dg W_{+1} + (\l_{+21}-S_{21})a^*_{+2}W_{+2}^\dg Z_1 + \nn \\
& \qquad (\l^*_{-12}-S_{21})a_{-1}Z_2^\dg W_{-1} + (\l_{-21}-S_{21})a^*_{-2}W_{-2}^\dg Z_1 \= 0 \label{d2}
\end{align}
Like before one obvious choice is $\l_{+12}=\l_{-12}=S_{12}$ which is trivial since each of $W_+$ and $W_-$ must fall in the earlier discussed case, but unlike before there is an alternative that is not equally trivial. If $Z_2 W_{+1}$ and $W_{-2} Z_1$ contain the same operators and the same holds for $Z_2^\dg W_{+1} $ and $W_{-2}^\dg Z_1$ (and similarly for the other pairs) then $W_{+,p}$ and $W_{-,p}$ can be written in the form
\eq{W_{+,p}=Z_p \Wt \qquad \text{and} \qquad W_{-,p}=Z_p \Wt^\dg}
where $\Wt$ has momentum zero and must be independent of $p$. If we define
\eq{Z_p \Wt \= \mu_p \Wt Z_p}
then $\l_{+12} = \mu_2^* S_{12} $, $\l_{-12} = \mu_2 S_{12}$ and the above conditions are equivalent to the set of equations 
\begin{align}
& \frac{a_{-2}}{a_{-1}} = \frac{\mu_2 -1}{\mu_1 -1} \\
& \frac{a_{+2}}{a_{+1}} = \frac{\mu_2^* -1}{\mu_1^* -1} \\
& \frac{a^*_{-2}}{a_{+1}} = -\frac{\mu_2^* -1}{\mu_1^* -1} \\
& \frac{a^*_{+2}}{a_{-1}} = -\frac{\mu_2 -1}{\mu_1 -1}
\end{align}
which have to be valid for any choice of $p_1,p_2$. The solution is
\begin{align}
& a_{+,p} = b (\mu^*_p - 1) \\
& a_{-,p} = -b^* (\mu_p - 1)
\end{align}
for some constant $b$.

A simple choice for $\Wt$ is the product of two operators that create a pair of particles with opposite momenta $Z^\dg_q Z^\dg_{-q}$. Obviously a sum over all such pairs can be used, leading to the following expression for $W_p$
\eq{W_p \= \sum_q b_q (S_{p,q} S_{p,-q}-1) Z_p Z_q^\dg Z_{-q}^\dg + \sum_q b^*_q (S_{p,q} S_{p,-q}-1) Z_{-q} Z_q Z_p}
where by symmetry of the sums under $q \to -q$ we find that $b_q$ can be chosen to satisfy $b_{-q}=b_q S_{q,-q}$. Substituting into the general conditions (\ref{1.1}) and (\ref{2.1}) we find out that an additional lower order term proportional to $Z_{-p}^\dg$ is necessary in $W_p$ in order to satisfy the full-form conditions. Overall the tranformation is
\eq[ex2b]{W_p = \sum_q b_q (S_{p,q} S_{p,-q}-1) Z_p Z_q^\dg Z_{-q}^\dg + \sum_q b^*_q (S_{p,q} S_{p,-q}-1) Z_{-q} Z_q Z_p + 2 b_{-p} Z^\dg_{-p}}
Note that the last term cannot be absorbed by reordering the operators of the first one.

Other choices for $\Wt$ are still possible and may lead to more complex families of transformations. One may also consider linear combinations of more than two products and continue in a similar fashion.

\newpage
\bibliography{paper}

\end{document}